\documentclass[11pt,twoside]{article}
\usepackage{asp2004}
\usepackage{psfig}
\usepackage{epsf}
\usepackage{graphics}
\usepackage{lscape}
\def\edcomment#1{\iffalse\marginpar{\raggedright\sl#1\/}\else\relax\fi}
\reversemarginpar

\begin{document}
\title{Supernova Asymmetries and Pulsar Kicks ---\\ 
Views on Controversial Issues}
\author{H.-Th. Janka$^1$, L. Scheck$^1$, K. Kifonidis$^1$, 
E. M\"uller$^1$, and T. Plewa$^2$}
\affil{$^1$Max-Planck-Institut f\"ur Astrophysik, Karl-Schwarzschild-Str. 1,
D-85741 Garching, Germany}
\affil{$^2$Center for Astrophysical Thermonuclear Flames, University
of Chicago, 5640 S. Ellis Avenue, Chicago, Illinois 60637, USA}

\begin{abstract}
Two- and three-dimensional simulations 
dem\-onstrate that hydrodynamic instabilities can lead to low-mode 
($l=1,\,2$) asymmetries of the fluid flow in the neutrino-heated layer 
behind the supernova shock. This
provides a natural explanation for aspherical mass ejection and for
pulsar recoil velocities even in excess of 1000 km/s. We propose
that the bimodality of the pulsar velocity
distribution might be a consequence of a dominant $l=1$ mode
in case of the fast component, while higher-mode anisotropy 
characterizes the postshock flow and SN ejecta during the birth of 
the slow neutron stars.
We argue that the observed large asymmetries of supernovae and 
the measured high velocities of young pulsars therefore do not 
imply rapid rotation of the iron core of the progenitor
star, nor do they require strong magnetic fields to
play a crucial role in the explosion. Anisotropic neutrino emission
from accretion
contributes to the neutron star acceleration on a minor level,
and pulsar kicks do not make a good case for non-standard neutrino
physics in the nascent neutron star.
\end{abstract}
\thispagestyle{plain}

\section{Introduction}


In the past years much work has been devoted to start
exploration of the very wide and certainly interesting 
parameter space associated with rapid rotation and,
linked to it, with the growth of strong magnetic fields in 
stellar core collapse and supernova (SN) explosions. These
studies have different motivation. Some of them aim at 
computing templates of gravitational-wave signals for the now 
operational interferometer experiments 
(e.g., Dimmelmeier, Font, \& M\"uller 2002a,b; Ott et al.\ 2004;
Kotake, Yamada, \& Sato 2003a). Some of them intend to study 
the differences between neutrino-driven 
SN explosions of non-rotating progenitors and only 
recently available pre-collapse models with rotation (e.g., Fryer 
\& Heger 2000, Fryer \& Warren 2004, Kotake, Yamada, \& Sato 2003b).
Others are undertaken to support the idea that
magnetic fields provide the driving force of massive star
explosions and could generate MHD jets in SNe (e.g.,
Akiyama et al.\ 2003; Thompson, Quataert, \& Burrows 2004, 
Kotake et al.\ 2004, Obergaulinger 2004), a hypothesis which is 
inspired by the discovery of gamma-ray burst jets from 
collapsing stars, by polarization measurements and 
observed asymmetries of SNe, 
and by the growing evidence of highly-magnetised neutron
stars (NSs), the magnetars.

Rotation of the stellar iron core is considered here as
``rapid'' if it noticably affects gravitational collapse, 
core bounce, and early post-bounce evolution of the
SN. This requires {\em pre-collapse} rotation 
rates of significantly more than 1$\,$rad$\,$s$^{-1}$ in
the stellar center;
SN modelers typically assume values of 
3--10$\,$rad$\,$s$^{-1}$ or more to study rotational effects
during core collapse. A NS with 10$\,$km final radius will
spin with a period around 1$\,$ms if it forms from an
iron core that rotates rigidly with only 
$\sim 0.5\,$rad$\,$s$^{-1}$, provided angular momentum is
conserved during the formation process. Rapid differential
rotation after collapse must also be expected 
to strongly amplify even small initial seed fields by winding 
or by the magneto-rotational instability, in which case magnetic
field effects could certainly not be ignored in discussions
of the explosion mechanism.

But what is the theoretical and observational basis for the 
assumption that SN cores are in ``rapid'' rotation, that
rotation determines the geometry of the explosion, and that 
strong magnetic fields are a crucial ingredient for 
understanding the
start of the explosion? Are rapid rotation and magnetic fields
needed to solve the long-standing problem how massive stars
explode, and to explain why SNe are deformed and why
pulsars have large velocities? In the following we shall argue
that there is currently no solid ground for such claims.

NS spin periods at birth, e.g. of the Crab pulsar, are 
estimated to be longer than 10--20$\,$ms (in some cases
hundreds of ms), provided the deceleration can be calculated
backward in time by using the magnetic dipole model. 
NSs like the ones in SN~1987A or in Cassiopeia A 
rotate at much less than the Crab rate, or must have a very
weak surface field. In both cases there is no
trace of the energy output from a bright, Crab-like pulsar.
The compact X-ray source at the center of Cas~A is
four orders of magnitude fainter than Crab, and again there is 
no sign of the energy output from the spin-down of an 
initially very fast rotator. So there is no direct information
for rapid rotation of the SN {\em core}
in the gaseous remnants of both explosions. Although the
ring system of SN~1987A confirms large angular momentum in
the outer layers of the exploded star --- possibly due to
a binary merger event some 10,000 years before the explosion
took place (Podsiadlowski 1992; Podsiadlowski, Joss, \& 
Rappaport 1990) --- this does not mean that the 
SN core at collapse had rotated rapidly. 
The apparent prolate deformation
of the ejecta of SN~1987A, which has been interpreted as a
signature of rapid rotation with an axis perpendicular to the plane
of the inner, bright ring, should be taken with caution. 
Dust formation is likely to have an important influence on the
observational appearance of the ejected gas (McCray 2004). The
sizable polarization of the light of many SNe and, in particular, 
the growth of the polarization with time and thus with deeper view 
into the more and more transparent ejecta, are also no unambiguous
evidence that strong rotation is the origin of 
the underlying deformation.
{\em Any} physical process which triggers the
explosion in a largely aspherical way and imposes a global
directionality on the mass ejection will also produce
polarization that grows towards the center.
Wispy filaments in the outer parts of the Cas~A 
remnant are looked at as relics of a ``jet'' and a
``counter-jet'', but these structures are Si and not Fe rich
(Hwang et al.\ 2004). If linked to the center of the SN they
were probably caused by bipolar outflow {\em after} the launch
of the explosion (possibly associated with late accretion by 
the NS; Janka et al.\ 2004a). This is supported by the
fact that the NS appears displaced from the geometrical center
of the reverse shock and of the ejecta knots almost perpendicular
to the line of ``jet'' and ``counter-jet''
(Thorstensen, Fesen, \& van den Bergh 2001, Gotthelf et al.\ 2001)
so that a connection between NS recoil and the jet acceleration
seems to be disfavored.

\begin{figure*}[t!]
\tabcolsep=0.5mm
\begin{tabular}{lcr}
   \epsfxsize=0.32\hsize \epsffile{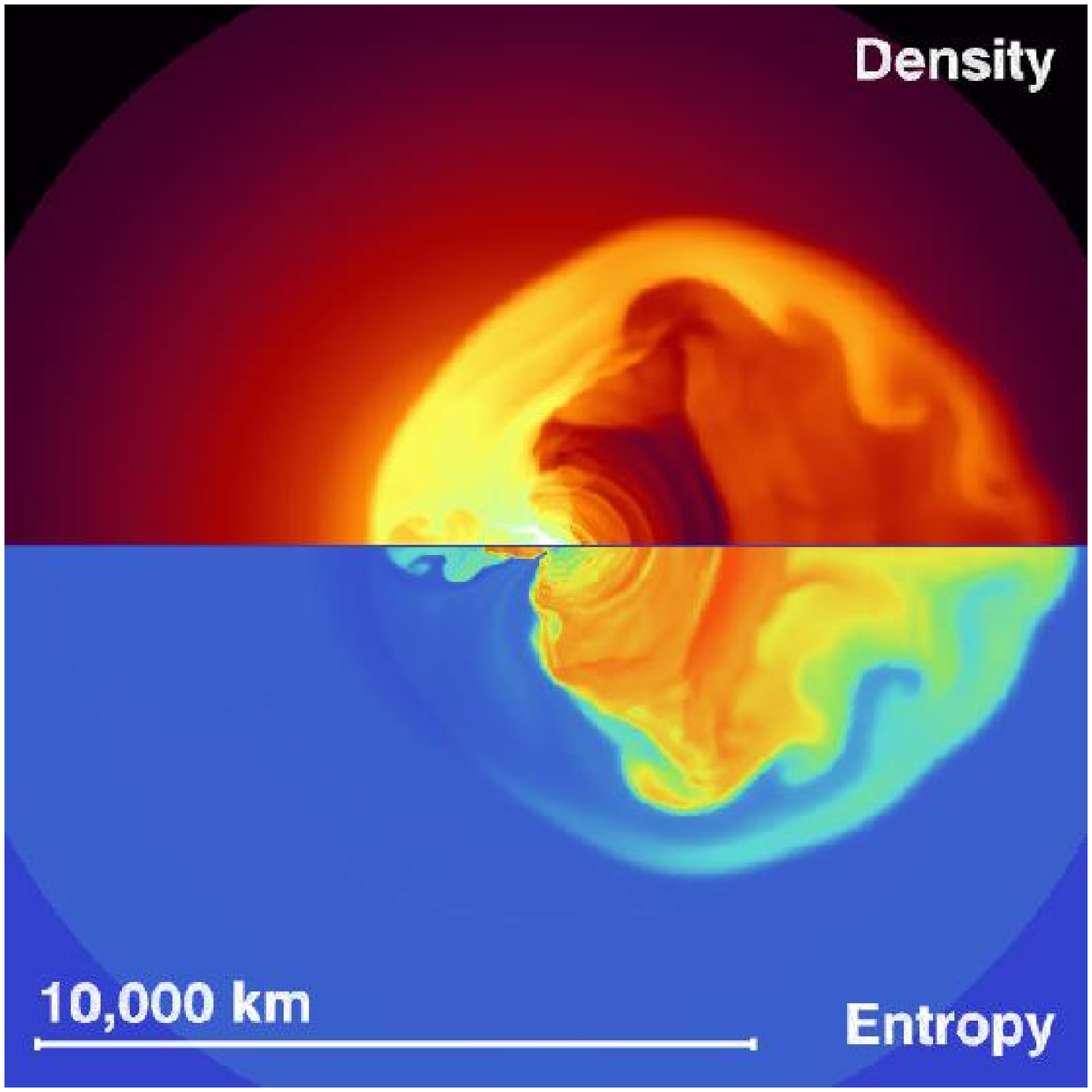} &
   \epsfxsize=0.32\hsize \epsffile{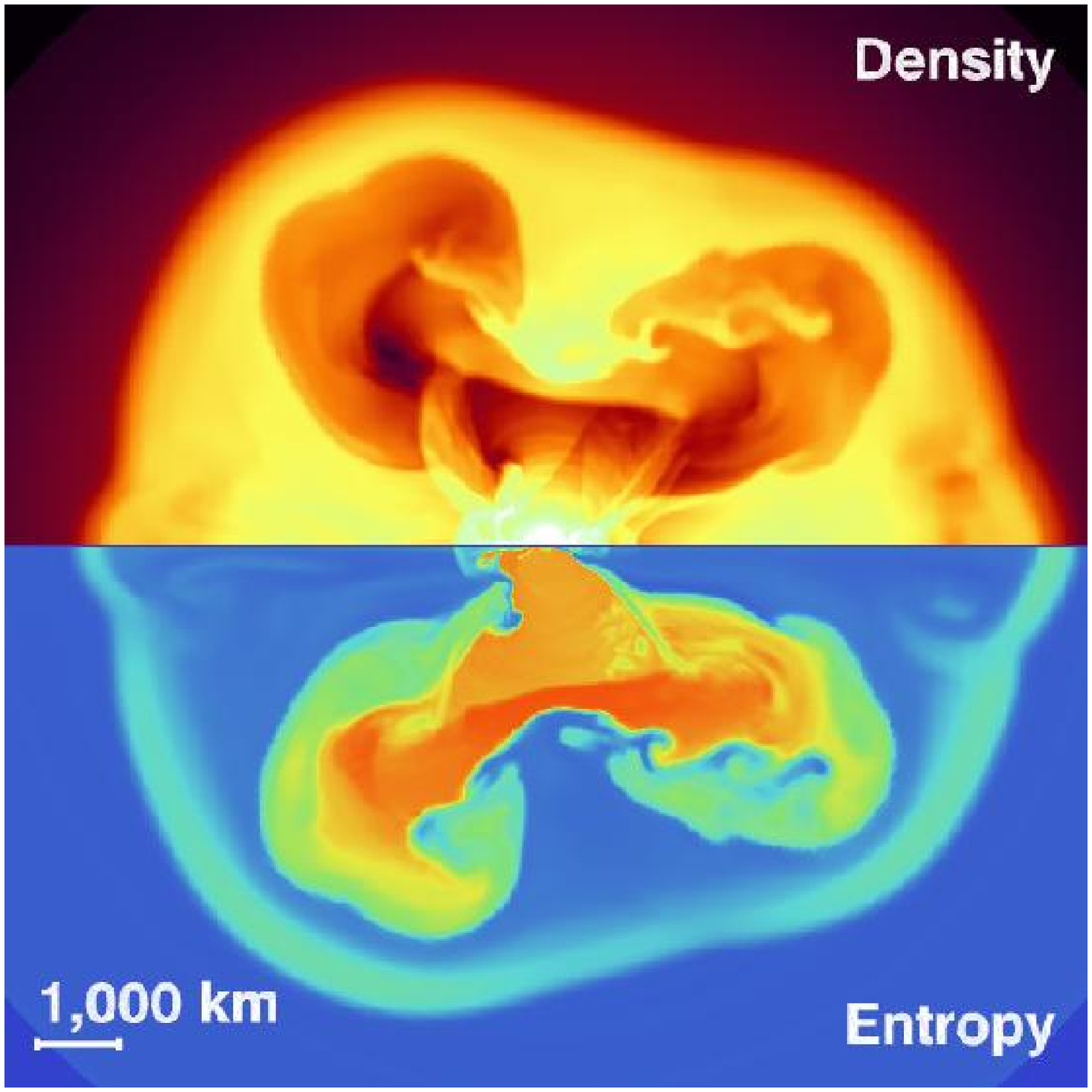} &
   \epsfxsize=0.32\hsize \epsffile{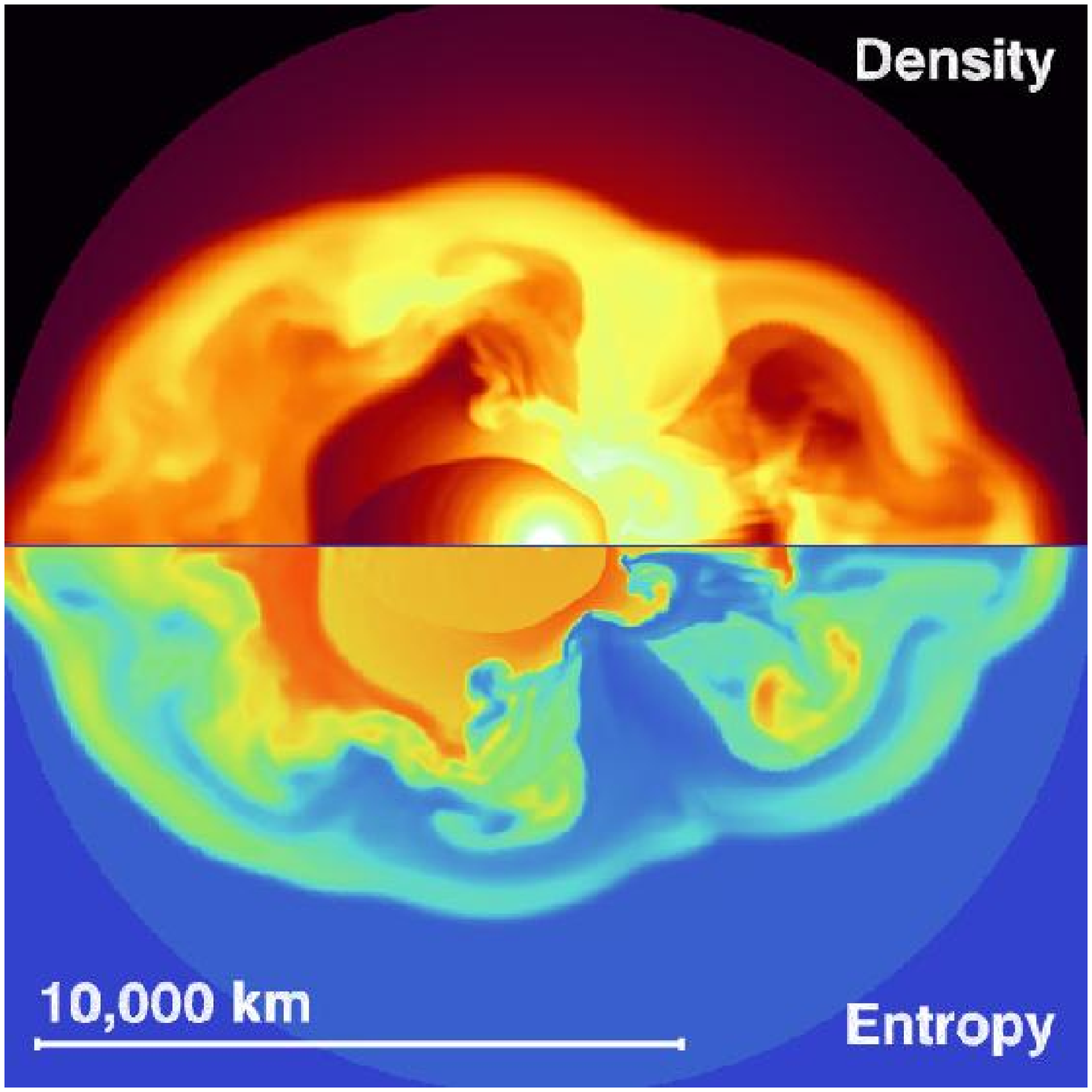} \\
   \epsfxsize=0.32\hsize \epsffile{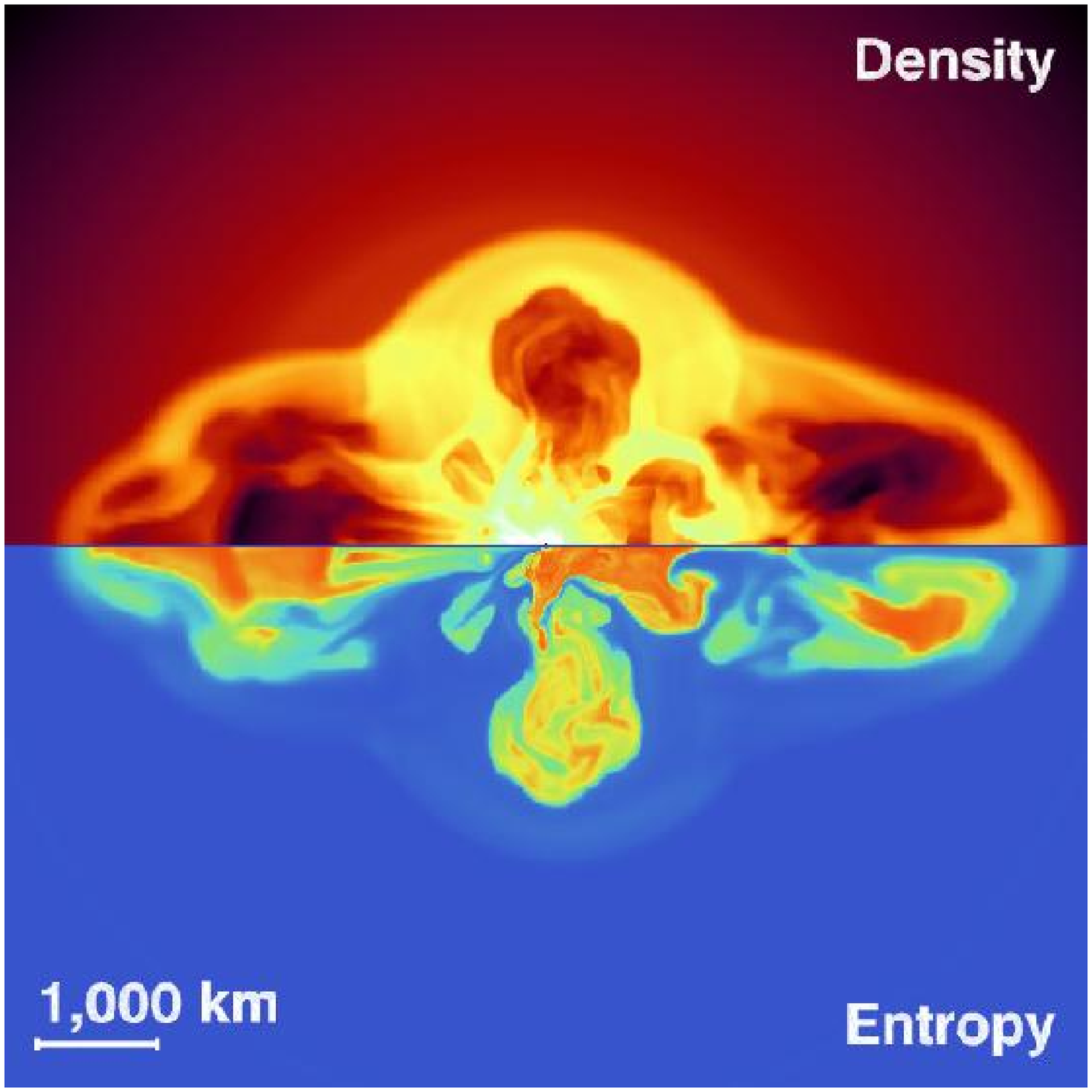} &
   \epsfxsize=0.32\hsize \epsffile{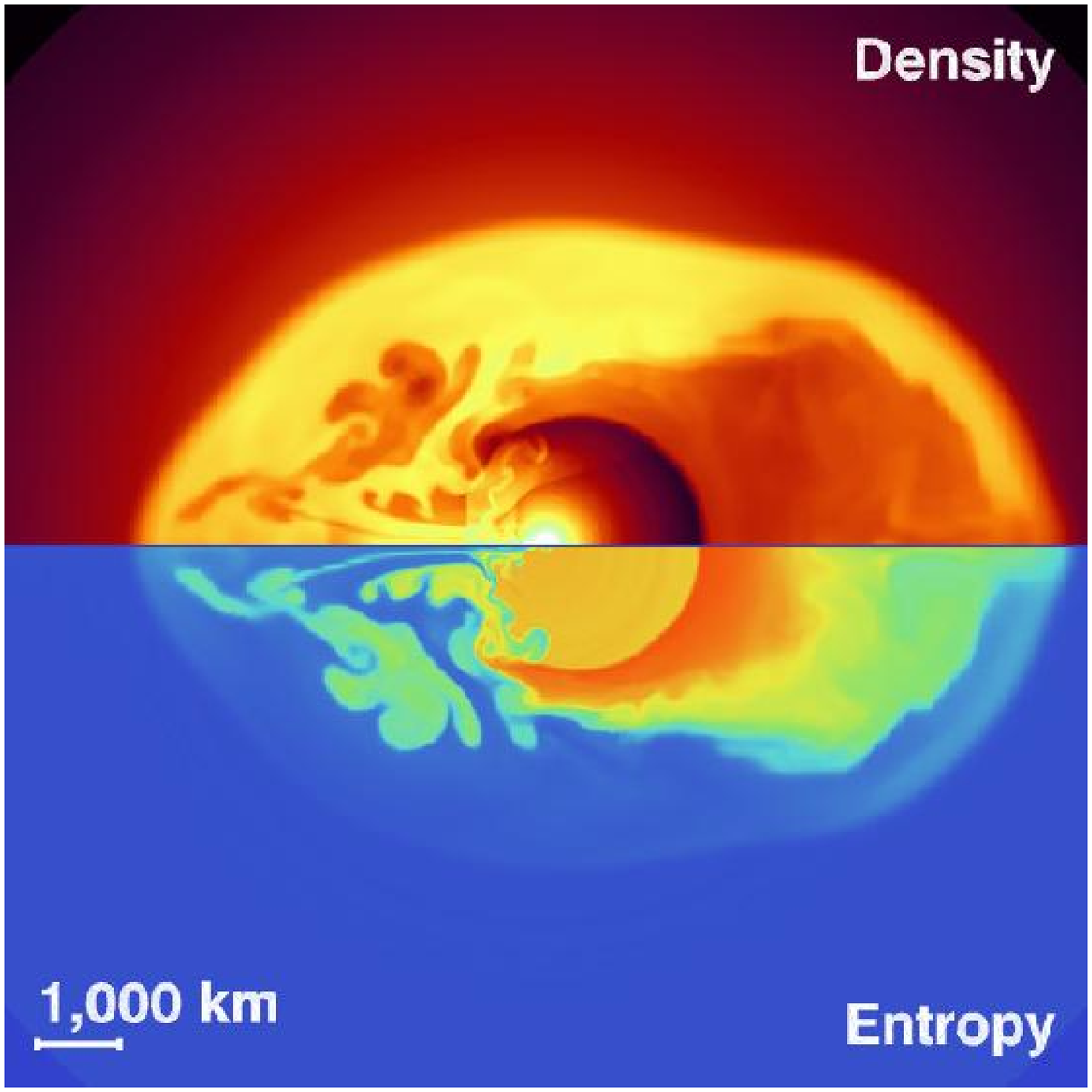} &
   \epsfxsize=0.32\hsize \epsffile{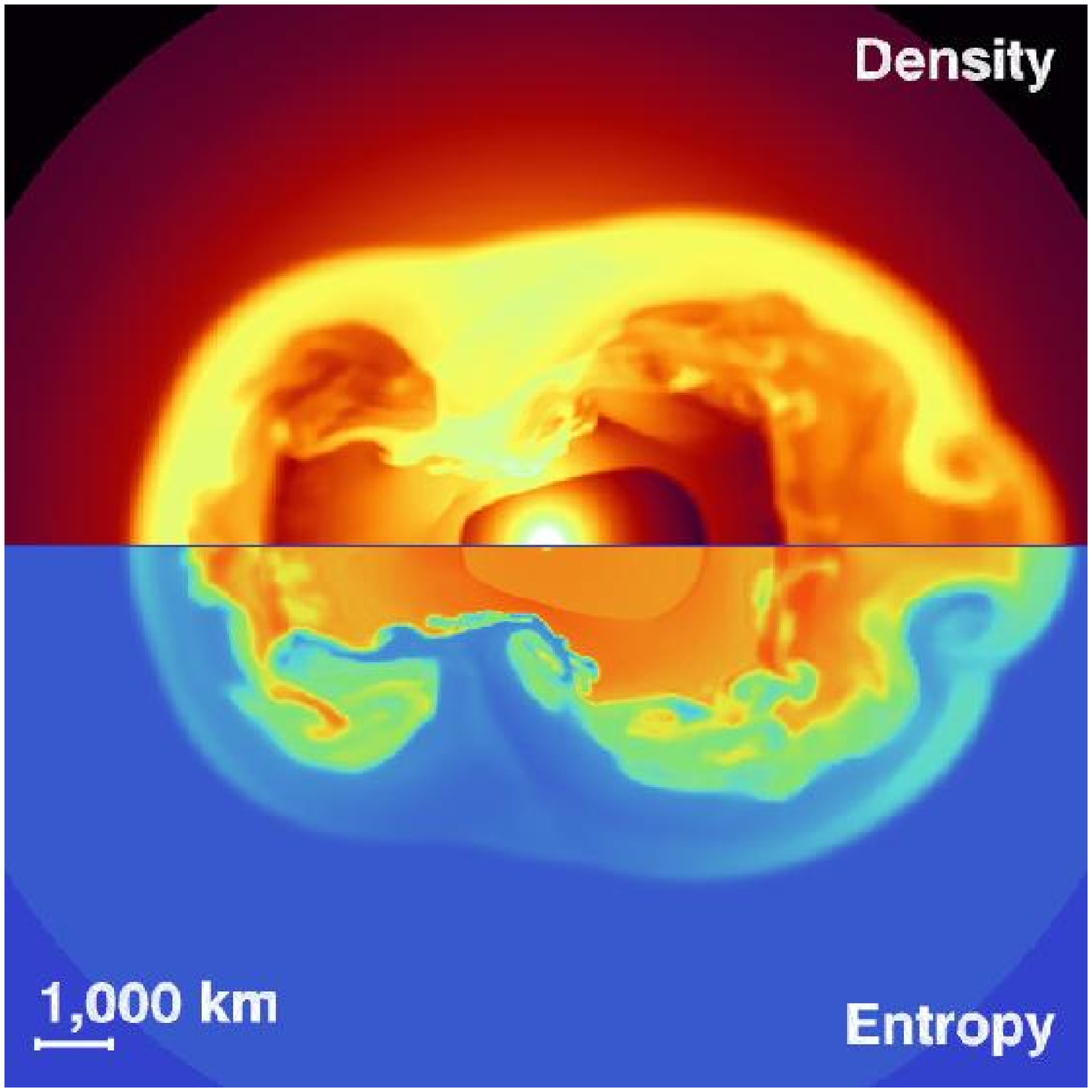}
\end{tabular}
   \parbox[t]{1.0\hsize}{\caption[]{
\small
Snapshots of six out of currently 66 2D supernova simulations at  
1$\,$s after bounce, showing different morphology as the result   
of the highly nonlinear growth of anisotropies from random seed
perturbations. The explosion energies at this time (still increasing
in some cases)
are 0.39, 0.96, 1.17, 0.33, 0.91, and $1.38\times 10^{51}$erg,
respectively (from top left to bottom right).
}
\label{janka_fig:snsample}}
\end{figure*}

Since massive stars suffer from mass loss prior to collapse,
they also lose significant amounts of angular momentum. 
Stellar evolution models including rotation 
(Heger, Langer, \& Woosley 2000,
Hirschi,  Meynet, \& Maeder 2004) show that convection and 
rotation-induced shear and circulation flows lead to efficient
transport of angular momentum out of the stellar core. This
is even enhanced when magnetic fields are taken into account.
Nonmagnetic stars develop iron cores with rotation rates
around 3--5$\,$rad$\,$s$^{-1}$ at the onset of collapse 
(Woosley, Heger, \& Weaver 2002), whereas the remaining angular
momentum is roughly 20 times smaller for magnetized cores 
(Heger et al.\ 2003a). While in the latter case the estimated spin
period of the newly formed NS is in fairly good agreement
with observations, the angular momentum in the 
collapsed stellar core in the former case exceeds the value
corresponding to the critical frequency of a compact NS. With angular
momentum being conserved during contraction, the NS would also gain 
a huge amount of rotational energy. For a 10$\,$km object with 
a period of 1$\,$ms the rotational energy is several $10^{52}$erg. 
This energy is neither measured in the SN explosion nor released
in the pulsar-powered remnant (calorimetry reveals the Crab Nebula, 
e.g., as the relic of a low-energetic SN), so would have to disappear 
via an invisible channel\footnote{Gravitational-wave 
emission by r-modes has been discussed as such a possibility, 
but the saturation amplitude turned out to be too low 
(Arras et al.\ 2003, Woosley \& Heger 2003).}. 
Alternatively, the rapidly rotating
stellar core could be decelerated right after collapse, {\em before} 
contraction to NS size has happened. But a process which 
transports angular
momentum sufficiently efficiently during this phase without 
braking the core rotation more efficiently during the much longer 
pre-collapse evolution, has not been identified.

In summary, it is unlikely that the cores of SN progenitors and
newly formed NSs rotate rapidly. The explosion 
mechanism and observational properties of ordinary SNe, e.g.,
their explosion energies, Ni nucleosynthesis, anisotropies, 
and pulsar kicks, therefore call for an explanation
that does not rely on the presence of large angular
momentum or on the rotational amplification of magnetic fields.
This is different from very energetic massive star explosions
(``hypernovae'') and gamma-ray burst events, where high 
rotation rates, jets, and possibly the formation of a black 
hole instead of a NS may be responsible for their particular
characteristics (Heger et al.\ 2003b).

\begin{figure*}[htp!]
\tabcolsep=0.5mm
\begin{tabular}{lr}
   \epsfxsize=0.49\hsize \epsffile{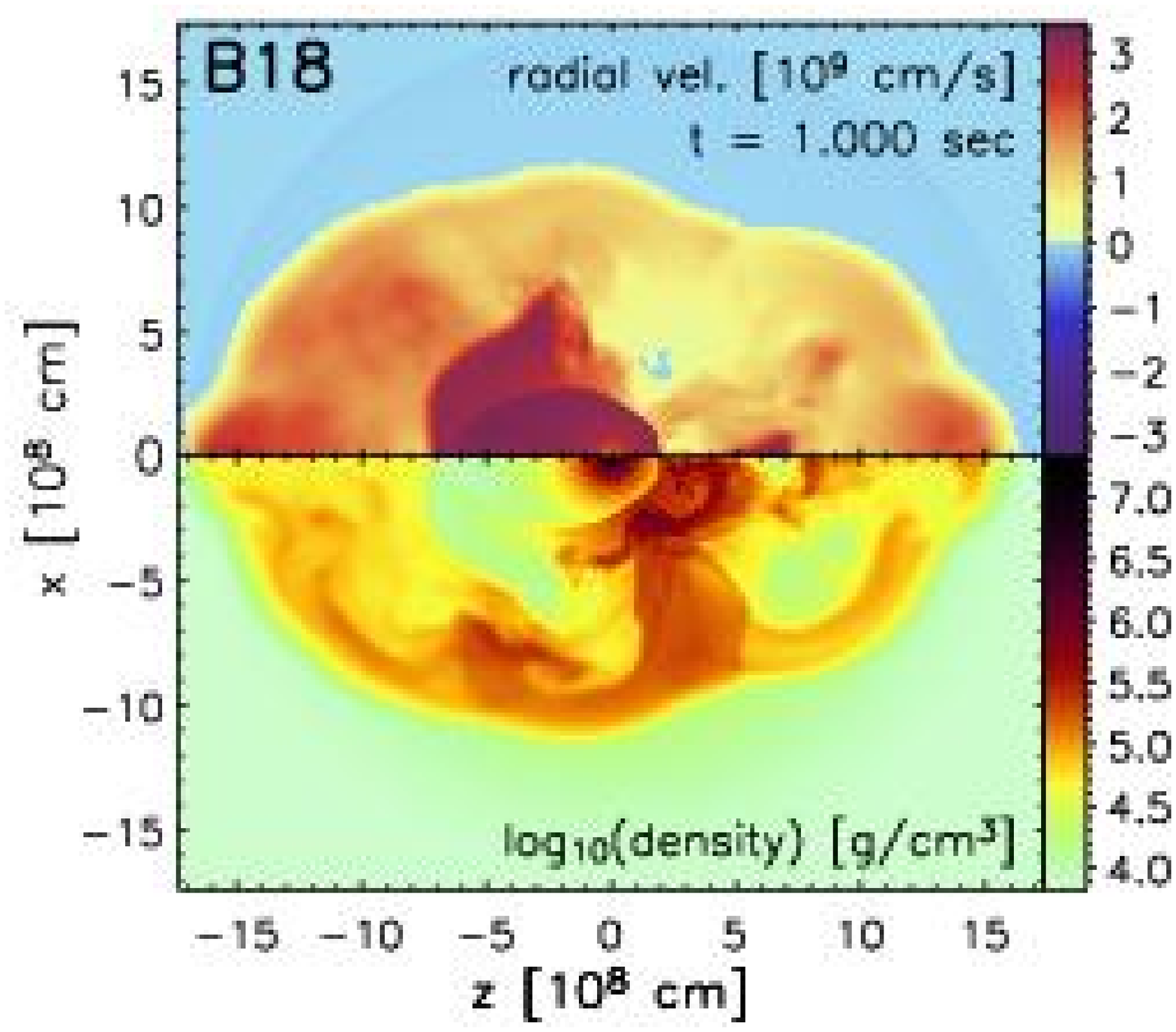} &
   \epsfxsize=0.49\hsize \epsffile{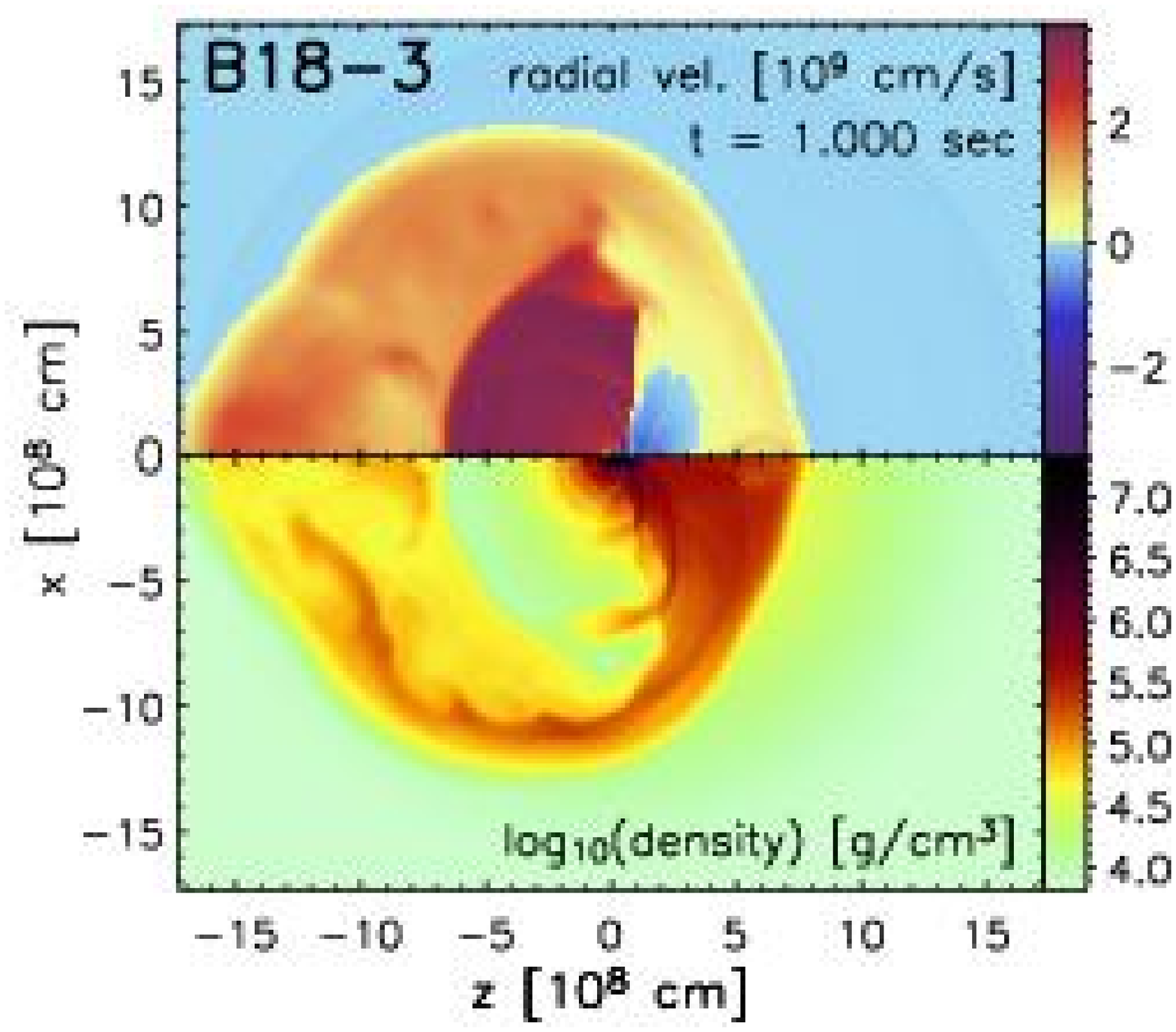} \\
   \epsfxsize=0.49\hsize \epsffile{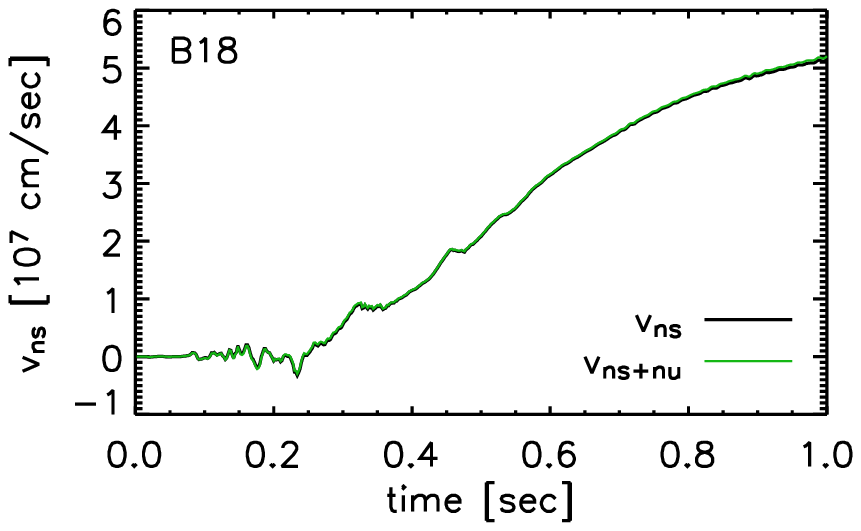} &
   \epsfxsize=0.49\hsize \epsffile{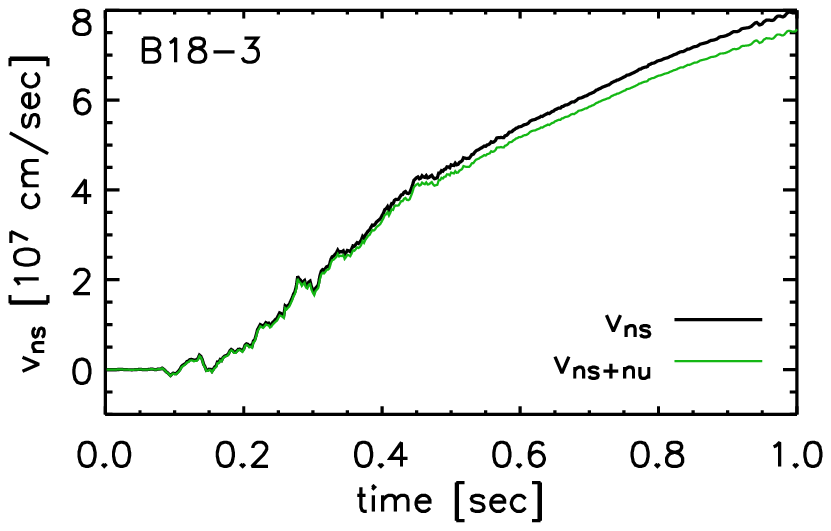} \\
   \epsfxsize=0.49\hsize \epsffile{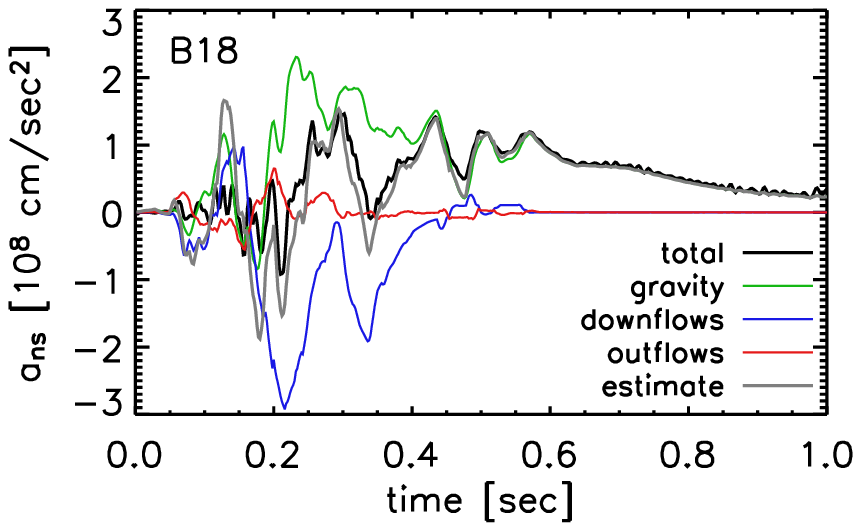} &
   \epsfxsize=0.49\hsize \epsffile{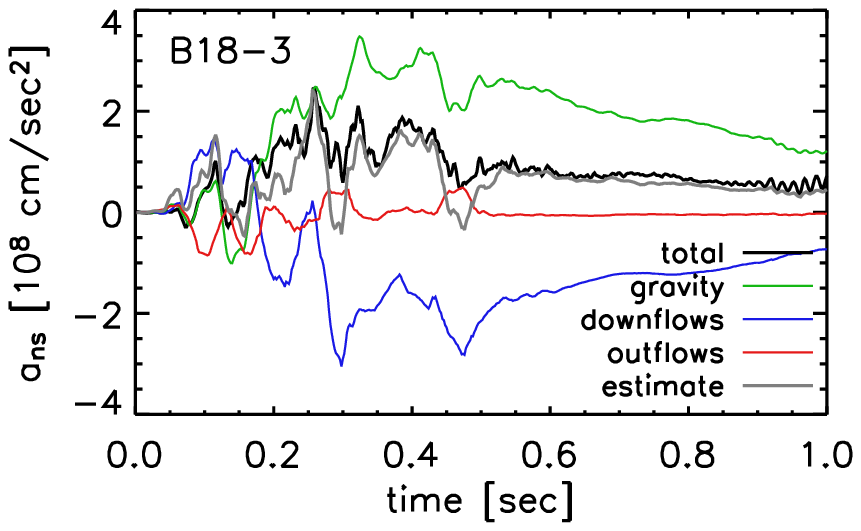}
\end{tabular}
   $\phantom{.}$
   \hskip -1.0truecm
   \parbox[t]{1.1\hsize}{\caption[]{
 \small
Two cases with large explosion asymmetry and dominant $l=1$
mode of the neutrino-heated ejecta, both yielding high NS recoil
velocities. The left model has a NS velocity of 520$\,$km/s,
the right model $\sim$800$\,$km/s after 1$\,$s post-bounce evolution.
In both cases the explosion energy is around $1.2\times 10^{51}\,$erg,
and the NS has still a very high acceleration at the end of our
simulations. The upper plots display the morphology of the ejecta
distribution at this time, the middle panels show the NS velocity
as a function of time, and the lower panels the corresponding
acceleration which is calculated from the negative of the rate of
momentum change of the SN ejecta (curve `total'). This net acceleration
agrees with the summed contributions (`estimate') from the
gravitational attraction between NS and surrounding gas (`gravity'),
momentum transfer to the NS by downflows (`downflows'), and recoil
by expanding, neutrino-heated gas (`outflows').
In the right model accretion continues until the simulation
was stopped, and gravity and accretion both contribute to the net
acceleration in the positive $z$-direction of the coordinate grid.
In contrast,
only gravitational forces provide the long-time acceleration of the
left model. Anisotropic emission of neutrinos from the accretion layer
as computed in our models makes only a minor effect for the NS
acceleration (i.e., the difference between the two lines in the 
middle panels).
}
\label{janka_fig:snasymm}} 
\end{figure*}

\section{Low-Mode Hydrodynamic Instabilities}

Large-scale deformation of the explosion and even a global asphericity,
however, do not require rapid rotation of the SN core but can be
caused by various kinds of hydrodynamic instabilities. Indeed we 
suspect that low-mode flow asymmetries may play a key role for 
explaining the observed inhomogeneities of the heavy-element 
distribution in SNe and SN remnants, the large
polarization measurements of SNe, and the high space velocities
of many young pulsars. 

Herant (1995) speculated about the possibility of 
a stable $l=1,\,m=0$ (one inflow, one outflow) convective mode
and discussed the potential importance of such a convective
pattern for NS kicks up to nearly 1000$\,$km/s. 
Herant's suggestion was motivated
by Chandrasekhar's (1981) finding that the easiest modes to excite 
in thermally unstable fluid spheres are those belonging to $l=1$. 
The situation discussed by Chandrasekhar resembles the one
developing in the SN core by the convective 
instability of the neutrino-heated region between gain radius
and SN shock, provided the radius of the latter is sufficiently 
much larger than the neutrinospheric radius and convection can
therefore become ``volume-filling''. Thompson (2000) also 
predicted instability of the accretion
shock to a global Rayleigh-Taylor mode 
that could lead to asymmetric shock expansion and a
net impulse of several 100$\,$km/s to the NS. Employing linear
stability analysis, Foglizzo (2001, 2002) identified highest
growth rates for non-radial $l=1$ instability of shocked accretion
flows due to the so-called ``entropic-acoustic cycle''
(Foglizzo \& Tagger 2000) in which the infall of entropy and
vorticity perturbations produces acoustic waves that 
propagate outward and create new entropy and vorticity 
perturbations when reaching the shock, thus closing an
amplifying feedback cycle. Blondin, Mezzacappa, \& DeMarino 
(2003) investigated the role of aspherical perturbations on
the stability of standing accretion shocks by idealized 
hydrodynamic models, and found 
that oblique shocks feed vorticity and entropy in the postshock 
region and lead to growing turbulence and shock instability
with an eventually dominant $l=1$ or $l=2$ mode.

\begin{figure*}[htp!]
\tabcolsep=3mm
\begin{tabular}{lcr}
   \epsfxsize=0.25\hsize \epsffile{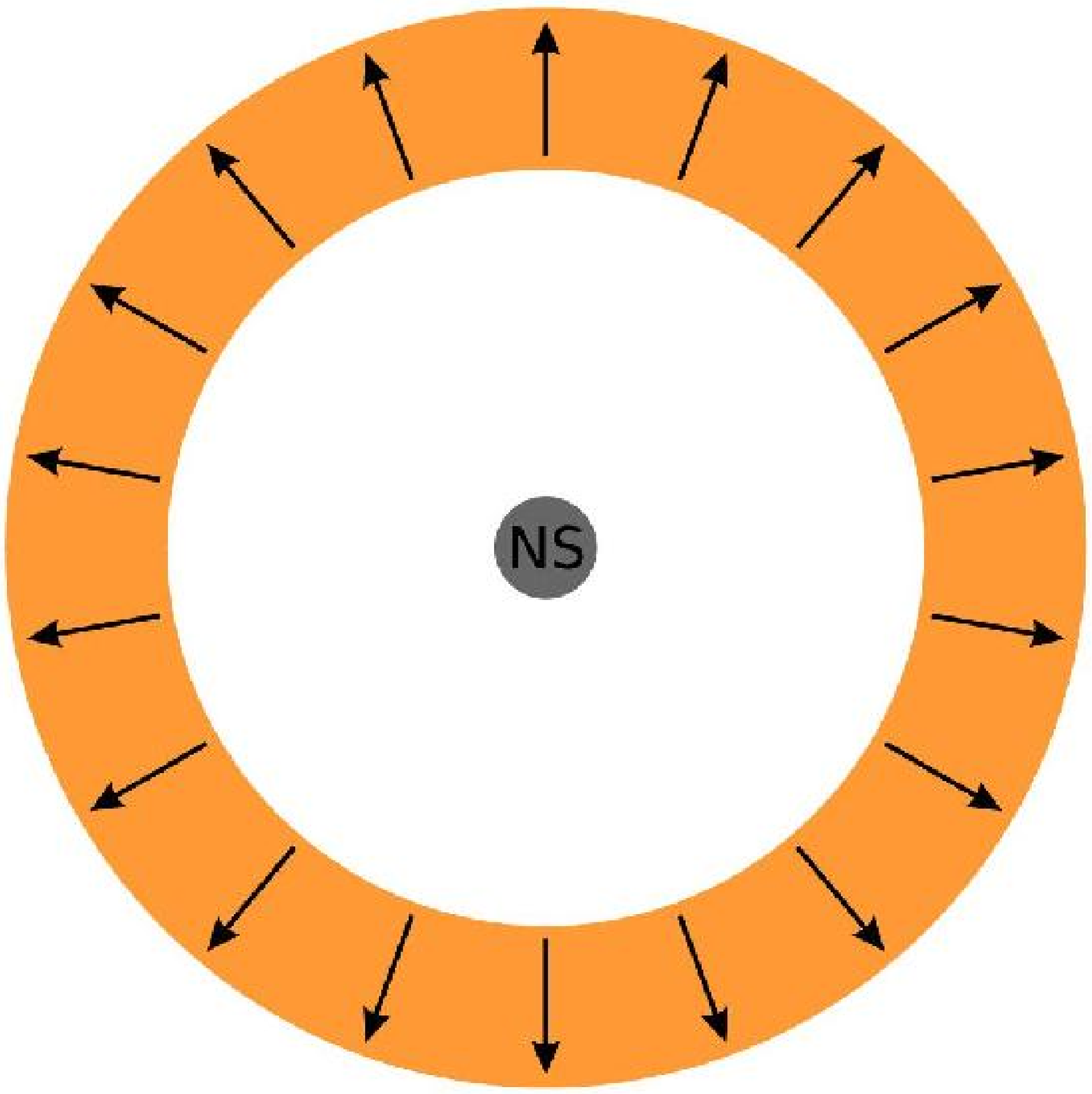}$\ \ \ \ \ $ &
   \epsfxsize=0.25\hsize \epsffile{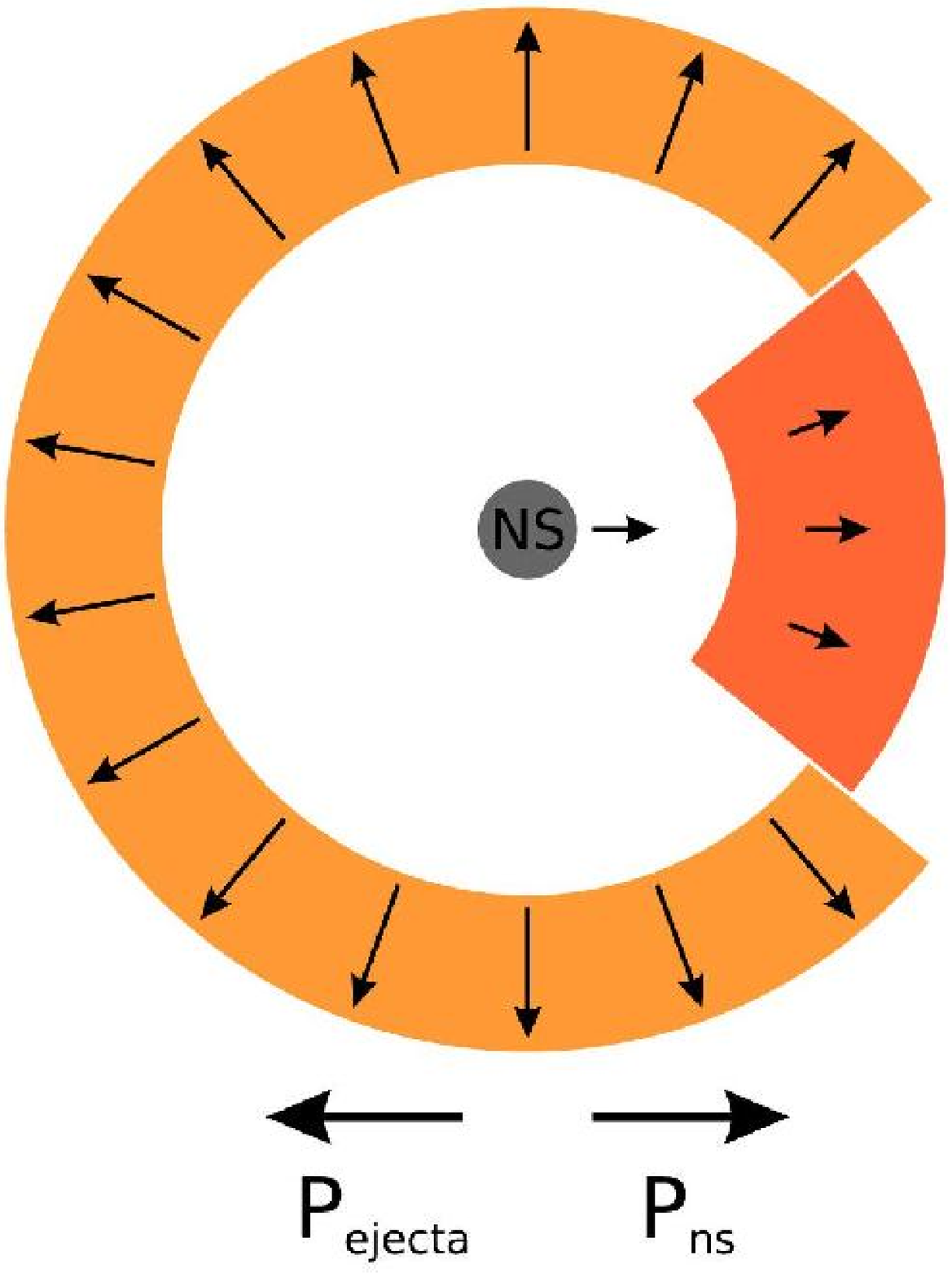}$\ \ \ \ \ $ &
   \epsfxsize=0.25\hsize \epsffile{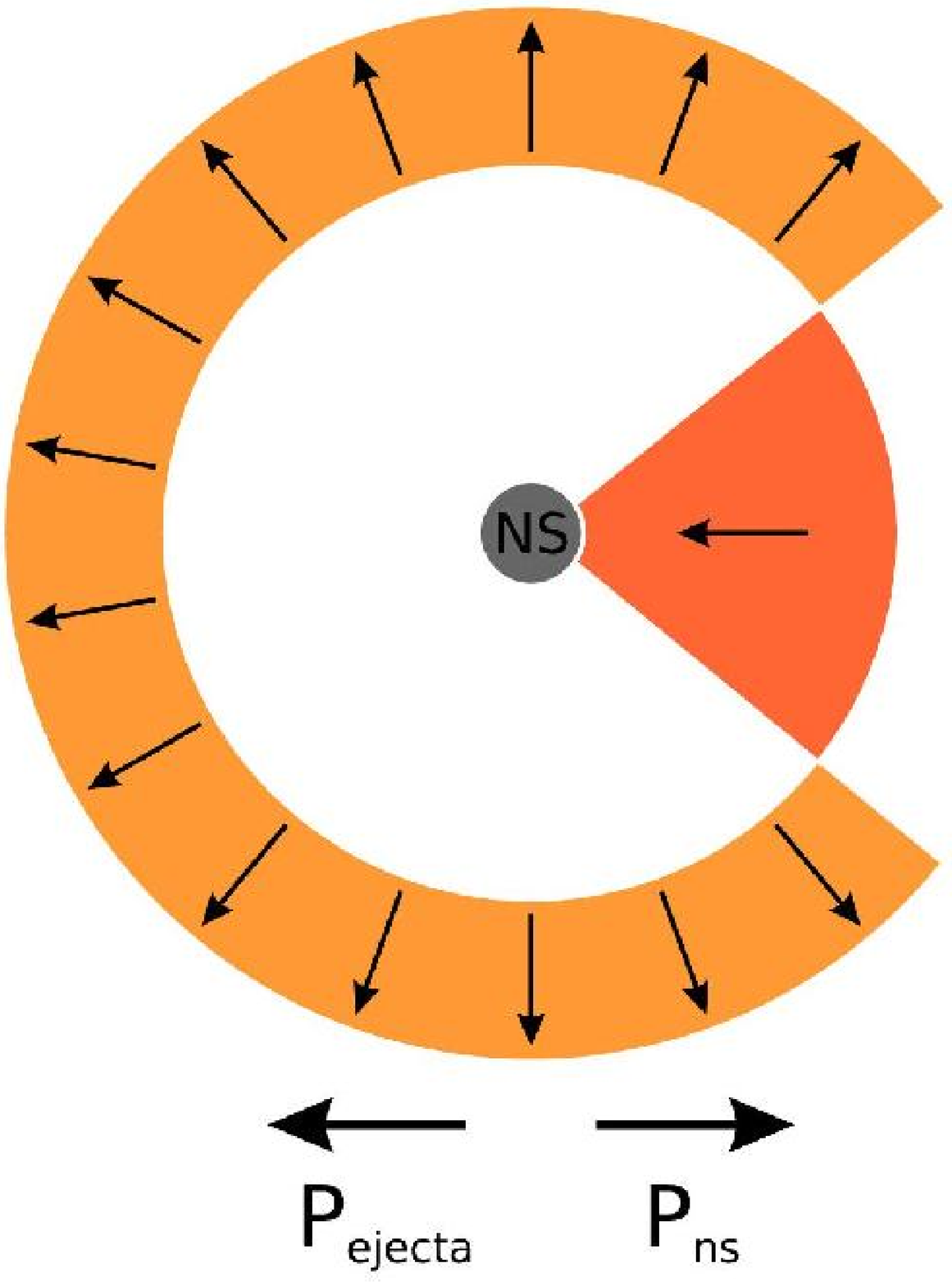}
\end{tabular}
  \parbox[t]{0.99\hsize}{\caption[]{
\small
Graphical illustration how the NS is accelerated
by a global asymmetry of the SN explosion. In case of a
spherically symmetric distribution of the ejecta the NS
remains at rest in the c.o.m. frame (left).
A recoil is obtained mainly
by the action of gravitational forces between NS and
anisotropic ejecta (middle; corresponding to the case in the left
panels of Fig.~\ref{janka_fig:snasymm}) or by the transfer of momentum
(hydrodynamic forces) in long-lasting accretion flows to the compact
remnant (right; cf.\ the model shown in the right panels of
Fig.~\ref{janka_fig:snasymm}).
The NS and ejected mass carry opposite momenta.
}
\label{janka_fig:acc}}
\end{figure*}

\begin{figure*}[t!]
\tabcolsep=0.5mm
\begin{tabular}{lr}
   \epsfxsize=0.66\hsize \epsffile{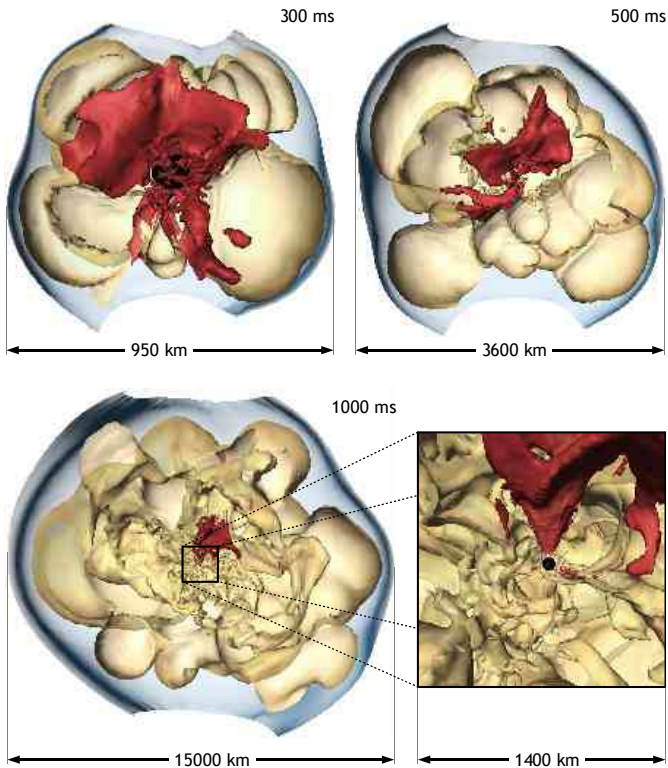} &
   \raisebox{10.5cm}{ 
   \parbox[t]{0.33\hsize}{\caption[]{
\small
Snapshots from a 3D simulation at 300$\,$ms (top left),
500$\,$ms (top right) and 1000$\,$ms (bottom) after bounce,
showing the evolution of the neutrino-heated bubbles
(visualized by surfaces of constant neutron-to-proton
ratio), shock (enveloping surface), and accretion flows to the
NS at the center (red iso-surfaces of the mass flow rate).
Accretion is still going on at 1$\,$s and has  
developed an $l=1$ mode (lower right enlargement).
}
\label{janka_fig:3d}}}
\end{tabular}
\end{figure*}

In fact, we have recently demonstrated the existence of such 
low-mode shock instabilities in the realistic SN environment by
2D hydrodynamic simulations with a detailed treatment of the 
equation of state and
neutrino ($\nu$) physics. The result did not depend on whether a
simplified $\nu$ transport with $\nu$
luminosities imposed at an inner, contracting grid boundary 
(supposed to mimic the $\nu$ emission from the shrinking high-density
core of the nascent NS) were used (Scheck et al.\ 2004),
or whether full-scale SN models were calculated employing
spectral (but still radial) transport by
a variable Eddington factor technique for solving the 
moment equations of lepton number, energy,
and momentum (Janka et al.\ 2004b,c). In both kinds of simulations
we could find a dynamical behavior much alike the one reported by
Blondin et al.\ (2003). When the explosion timescale is
sufficiently long and the shock radius relative to the NS
radius is sufficiently large, the convective cells in the 
$\nu$-heating layer have time and volume to merge to huge
structures (Fig.~\ref{janka_fig:snsample}).
We observed large non-radial shock oscillations
and the development of a bipolar deformation with axis ratios
up to more than 1:2. Finally the flow pattern behind the
expanding, aspherical SN shock is dominated by $l=1$ and
$l=2$ modes in self-similar expansion. The role of the 
vortical-acoustic instability in this process of mode merging 
is not obvious when violent convective activity is present.
But we directly observed this kind of non-radial shock 
instability in cases where the 
onset of postshock convection was suppressed by our choice of
a low $\nu$ luminosity from the inner boundary. With thus
reduced $\nu$ heating behind the shock the negative entropy
gradient was rather flat and high infall velocities
of the gas behind the accretion shock (standing at a
relatively small radius for these conditions) made the growth 
timescale of convection longer than the advection timescale.
Thus Rayleigh-Taylor instabilities could not sprout. Nevertheless,
the shock became unstable in a characteristic way by amplifying
sound wave and vortex activity as envisioned
by Foglizzo (2001, 2002). 

\section{Pulsar Kicks}

\begin{figure*}[t!]
\tabcolsep=0.5mm
\begin{tabular}{lr}
   \psfig{file=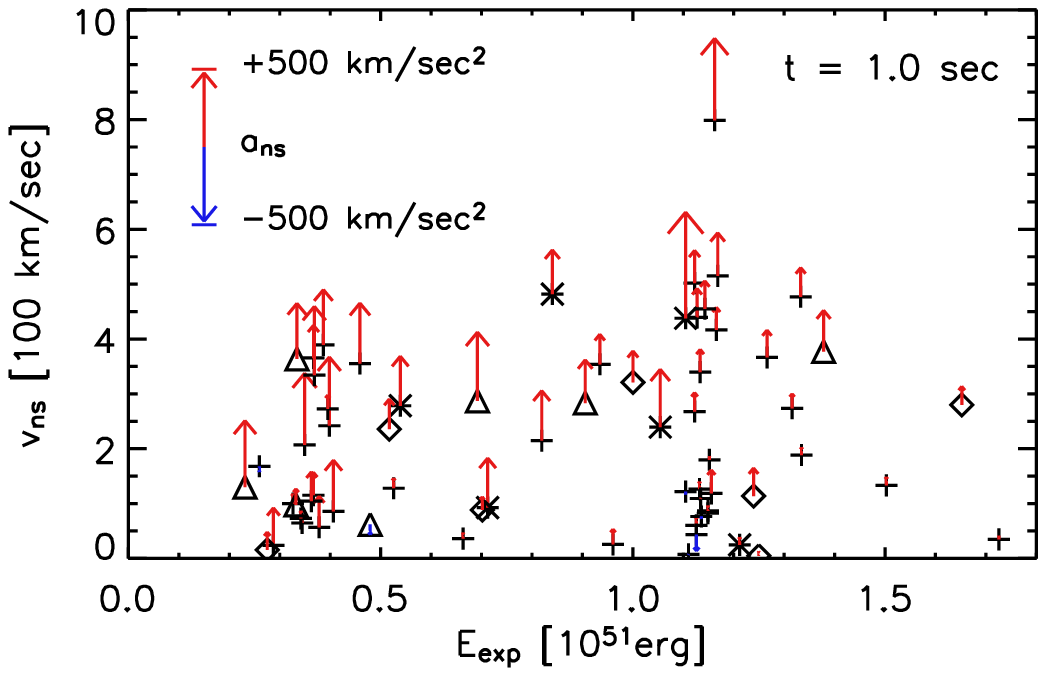,width=0.49\textwidth} &
   \psfig{file=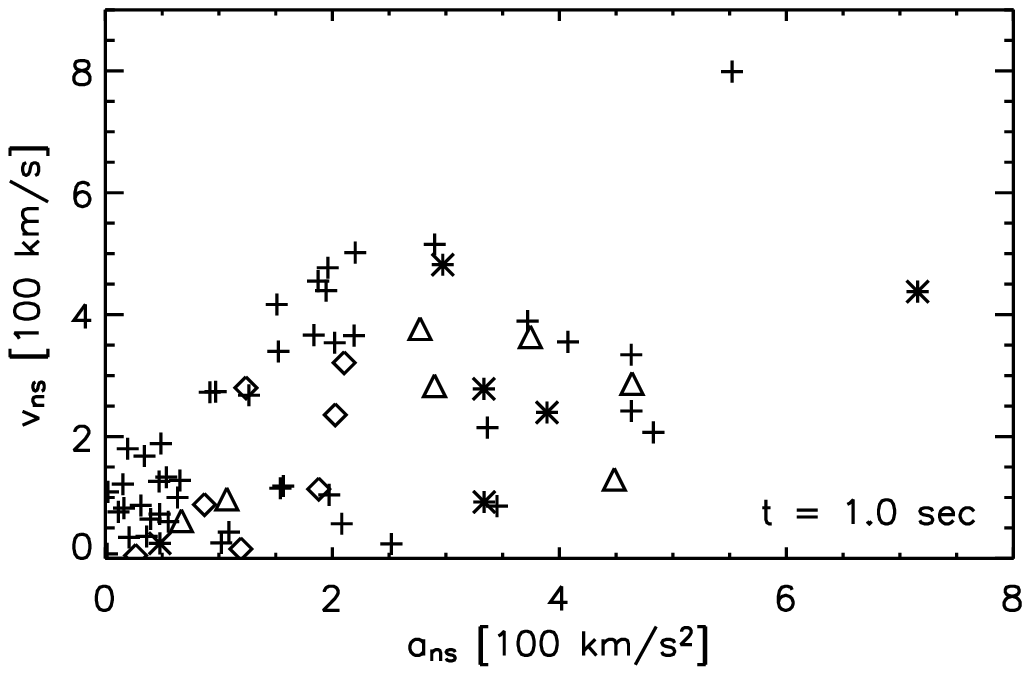,width=0.49\textwidth} \\
   \psfig{file=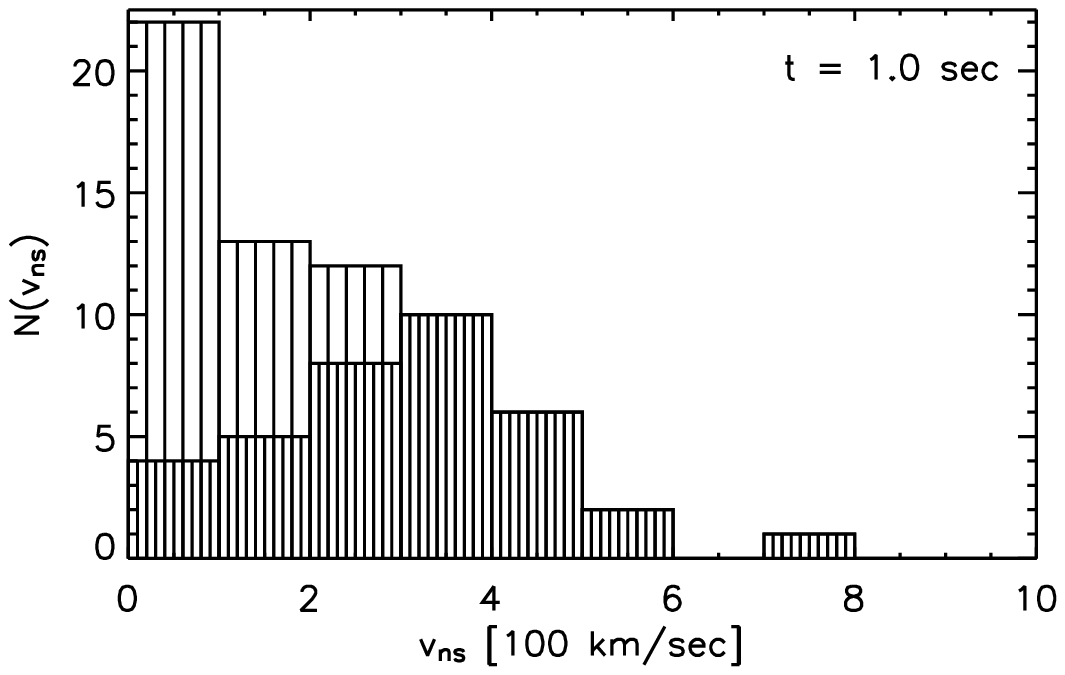,width=0.49\textwidth} &
   \psfig{file=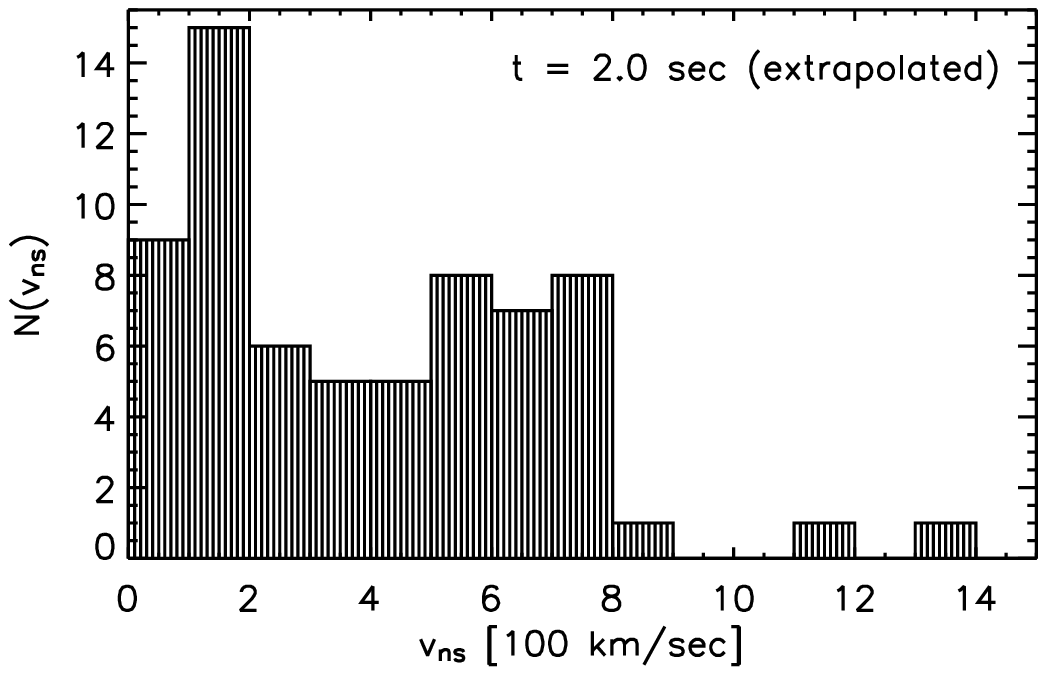,width=0.49\textwidth}
\end{tabular}
   \parbox[t]{0.99\hsize}{\caption[]{
\small
Statistics of currently 66 two-dimensional SN simulations
which were carried out to 1$\,$s after core bounce.
The upper left plot shows the NS velocity, $v_{\mathrm{ns}}$, 
(acceleration indicated
by arrows) vs.\ explosion energy $E_{\mathrm{exp}}$ and the upper right
plot $v_{\mathrm{ns}}$ vs.\ the NS acceleration, $a_{\mathrm{ns}}$, 
at 1$\,$s.
Different symbols correspond to different
$15\,M_{\odot}$ progenitors (see Scheck et al.\ 2004). The lower
left panel displays the number of cases vs.\ NS velocity at the end
of the simulations, binned in intervals of 100$\,$km/s, with
the dark-hatched area indicating those cases where $a_{\mathrm{ns}}$
is still larger than 150$\,$km/s$^2$ after 1$\,$s. Low-velocity 
neutron stars with small acceleration on the one hand and high-velocity
stars with large acceleration on the other (see also the upper left plot)
can lead to a bimodality 
which becomes visible when acceleration with the values at 1$\,$s is
assumed to continue for another second (lower right).
}
\label{janka_fig:statistics}}
\end{figure*}

In a large set of now 66 2D simulations for different
15$\,M_{\odot}$ progenitors, with and without rotation, 
Scheck et al.\ (2004) found that the explosion energy increases 
with higher core $\nu$ luminosity, but the kick velocity 
imparted to the newly formed NS varies stochastically with
the imposed initial nonradial perturbations 
(random seed with 0.1\% amplitude on velocity) in the SN core
(Figs.~\ref{janka_fig:snsample}, \ref{janka_fig:statistics}).
The anisotropy parameter $\alpha\equiv
|\int {\mathrm{d}}m\, v_z|/\int {\mathrm{d}}m\, |v_{\mathrm{gas}}|$
decreases with higher explosion energy, 
$E_{\mathrm{exp}}$, and thus with faster onset
of the explosion (i.e., less time for low-mode growth). But
since the net ejecta momentum $p_{\mathrm{ej}}\equiv 
\int {\mathrm{d}}m\, |v_{\mathrm{gas}}| \equiv 
M_{\mathrm{ej}}\left\langle |\vec v_{\mathrm{ej}}| \right\rangle$
increases essentially linearly with $E_{\mathrm{exp}}$, the 
kick velocity 
$v_{\mathrm{ns}} = \alpha\, p_{\mathrm{ej}}/M_{\mathrm{ns}}$
shows no obvious correlation with $E_{\mathrm{exp}}$
(Fig.~\ref{janka_fig:statistics}, upper left panel).

After 1$\,$s of post-bounce evolution, we obtained NS velocities
up to 800$\,$km$\,$s$^{-1}$ (to our knowledge this is the 
world record of {\em simulated} kicks), 
in many cases associated with still large acceleration so that 
final velocities well above 1000$\,$km$\,$s$^{-1}$ can be expected
(Fig.~\ref{janka_fig:statistics}, upper two panels). 
Figure~\ref{janka_fig:snasymm} shows the record holder (right) and
another high-velocity case, both having positive acceleration
at the end of our simulations due to a combination of momentum 
transfer in case of ongoing accretion and the gravitational
attraction of anisotropically distributed ejecta 
(Fig.~\ref{janka_fig:acc}). Anisotropic $\nu$ emission by 
accretion, which is included in our simulations, turned out to
contribute to the NS acceleration only on a minor level
(Fig.~\ref{janka_fig:snasymm}). The NS is kept fixed at the grid
center because its motion is inhibited by the use of the inner boundary.
In order to test this constraint we changed 
the frame of reference in some models by applying
a coordinate transformation, 
adding a global, coherent acceleration on the whole grid with
the value of the NS acceleration and opposite to its direction.
Of course, the result for a particular choice of parameters changed, 
but no fundamental differences of the ensemble behavior were discovered. 
A first 3D simulation revealed also the 
development of an $l=1$ mode in the accretion flow to the nascent
NS within the first second of SN evolution (Fig.~\ref{janka_fig:3d}).

Although statements on the basis of our current sample of models
(limited to $15\,M_{\odot}$ progenitors; 2D models
instead of 3D; use of imposed core 
$\nu$ flux and simplified transport instead of fully self-consistent
explosions; binary effects ignored) have to be made with caution, 
we propose here a speculative possibility for the origin of the
bimodality of the observed velocity distribution (e.g., 
Arzoumanian, Chernoff, \& Cordes 2002). In
Fig.~\ref{janka_fig:statistics} one can see
an indication of two populations in our sample: One big group 
(in the lower left corner of the upper right panel) has {\em low
velocities and low acceleration} at 1$\,$s; these models have not
produced a dominant $l=1$ mode in the ejecta. The second big
group (towards the upper right in the same panel) has {\em high 
velocities and high acceleration} at 1$\,$s due to the dominance
of the $l=1$ asymmetry. This situation is also visible in the
lower left panel. The lower right panel of 
Fig.~\ref{janka_fig:statistics} shows the distribution 
extrapolated to a time
2$\,$s after core bounce, assuming that the acceleration continues
with a constant value between 1$\,$s and 2$\,$s. A bimodality 
becomes visible which appears very similar to the bold solid
line in Fig.~3 of the Arzoumanian et al. (2002) paper. The 
minimum of the distribution between the low velocity 
($v_{\mathrm{ns}} \la 200\,$km$\,$s$^{-1}$) and high velocity
($v_{\mathrm{ns}} \ga 300\,$km$\,$s$^{-1}$) components develops 
only on a timescale of possibly many seconds, because
the large acceleration by an $l=1$ asymmetry continues much
beyond the first second after bounce.

\section{Conclusions}

We have presented results which demonstrate that hydrodynamic 
instabilities of the stalled accretion shock and the $\nu$-heated 
postshock layer can lead to global anisotropy of the ejecta momentum 
and energy by the
dominance of $l=1,\,2$ modes. These asymmetries do
neither require rapid rotation nor the presence of strong
magnetic fields in the SN core. They generically occur during the
post-bounce accretion phase of the stalled SN shock provided the
conditions during this phase are suitable.
The $\nu$-heating mechanism 
in combination with such low-mode hydrodynamic instabilities thus
seems to yield a unique and consistent explanation for the
SN explosion on the one hand, and for
the observed asymmetries of SNe and the measured
space velocities of pulsars including
the bimodality of the NS velocity distribution on the other.

If our suggestion is valid (and, admittedly, the possible 
objections are still many), pulsar kicks would be a consequence  
of explosion asymmetries. In this case the SN ejecta and the NS
would carry equal linear momenta in opposite directions, a fact 
that might allow for a verification by future observations. 
This would alleviate
the need to invoke anisotropic $\nu$ emission from the nascent
NS as an explanation for the pulsar kicks. Although already 
$\sim$1\% asymmetry is sufficient to account for 
$300\,$km$\,$s$^{-1}$, it is extremely difficult to produce
emission anisotropies even at this low level, a fact which
instigated claims for the presence of very strong 
magnetic fields and/or non-standard $\nu$ physics (for a review,
see Lai, Chernoff, \& Cordes 2001).
Magnetars are often quoted in support of very high field strengths
in NSs. However, there is no observational hint for a correlation
of large fields and large pulsar velocities. Pulsar recoil by
$\nu$ oscillations (Kusenko \& Segre 1996), for example, requires 
the presence of fields with a very strong dipole component but 
also the existence of a sterile $\nu$ (Fuller et al.\ 2003 
and references therein), because resonant flavor conversions of 
active $\nu$'s occur at densities below 10$^4$g$\,$cm$^{-3}$
far outside of the NS for the $\nu$ mass differences 
of solar and atmospheric oscillations. The fields for kicks of
$300\,$km$\,$s$^{-1}$ are estimated to be huge, typically 
$\ga 10^{16}\,$G. The actual numbers are likely to be 
even higher (Janka \& Raffelt 1998), because all 
estimates are based on adhoc assumptions about the asymmetry of 
the geometry without taking into account that any anisotropy
of the $\nu$ emission unavoidably leads to an adjustment of the
structure of the NS. The corresponding feedback 
typically reduces the emission asymmetry (Janka \& Raffelt 1998).
Meaningful estimates of the latter therefore require 
self-consistent models of structure and transport.

{\small
\acknowledgments
Support by the 
SFB-375
``Astro-Teil\-chen\-phy\-sik'' of the 
DFG
is acknowledged. We also thank the John von Neumann -- Institut
f\"ur Computing (NIC) in J\"ulich for computer time on the 
IBM p690-Cluster Jump.

}

\end{document}